\documentclass[twocolumn,english,notitlepage,superscriptaddress,nofootinbib, twocolumn]{revtex4-1}
\usepackage[T1]{fontenc}
\usepackage[latin9]{inputenc}
\usepackage{bm}
\usepackage{amsmath}
\usepackage{amssymb}
\usepackage{graphicx}
\usepackage{comment}
\usepackage{enumerate}

\makeatletter
\usepackage{babel}

\usepackage{hyperref}
\hypersetup{
    colorlinks=true,
    linkcolor=blue,
    citecolor=red,
    filecolor=blue,      
    urlcolor=blue
     }

\newcommand{\afffive}{School of Physics and Astronomy, University of Glasgow, Glasgow, G12 8QQ, United Kingdom.}

\newcommand{\affone}{University College London, Gower Street, WC1E 6BT London, United Kingdom.}

\newcommand{\affthree}{Van Swinderen Institute, University of Groningen, 9747 AG Groningen, The Netherlands.}

\makeatother

\usepackage{babel}
\begin{document}
\title{Loss of coherence and coherence protection from a graviton bath}
\author{Marko Toro\v{s}}
\affiliation{\afffive}
\author{Anupam Mazumdar}
\affiliation{\affthree}
\author{Sougato Bose}
\affiliation{\affone}
\begin{abstract}
We consider a quantum harmonic oscillator coupled with a graviton
bath and discuss the loss of coherence in the matter sector due to
the matter-graviton vertex interaction. Working in the quantum-field-theory framework,
we obtain a master equation by tracing away the gravitational field
at the leading order $\mathcal{\sim O}(G)$ and  $\sim\mathcal{O}(c^{-2})$. We find that the decoherence
rate is proportional to the cube of the harmonic trapping frequency
and vanishes for a free particle, as expected for a system without
a mass quadrupole. Furthermore, our quantum model of graviton emission
recovers the known classical formula for gravitational radiation from
a classical harmonic oscillator for coherent states with a large occupation number. In addition, we find that the quantum harmonic oscillator
eventually settles in a steady state with \emph{a remnant coherence}
of the ground and first excited states. While classical emission of gravitational waves would make the harmonic system loose all of its energy, our quantum field theory model does not allow the number states $\vert 1\rangle$ and $\vert 0\rangle$ to decay via graviton emission. In particular, the superposition of number states $\frac{1}{\sqrt{2}}\left[\vert0\rangle+\vert1\rangle\right]$ is a steady state and never decoheres. 
\end{abstract}
\maketitle

\section{Introduction}

One of the most striking consequences of General relativity is undoubtedly
given by gravitational waves~\cite{misner1973gravitation}. Such
waves propagate through spacetime itself -- are part of it -- and
interact with all matter making it a universal feature of all experiments.
The gravitational waves produced by small objects are however hindered
by the smallness of gravitational coupling, whilst the gravitational
waves produced by large astronomical bodies become attenuated by the
large distances to the Earth. Nevertheless, a hundred years from the
prediction of gravitational waves~\citep{einstein1916,einstein1918gravitationswellen}
the detection of gravitational waves was announced~\citep{abbott2016observation}.

The feeble strain induced by the passing of gravitational waves
has been detected in an optomechanical setup employing suspended mirrors~\citep{abbott2016observation}.
Whilst most quantum effects remain suppressed at such scales it has
been shown that tiny quantum correlations between the phase of light
and the position of the mirrors in the Advanced LIGO detectors imprint
a non-negligible signal~\citep{yu2020quantum}. Furthermore, there
is substantial progress towards reaching the motional ground state
of $\sim10\text{kg}$ large mirrors where quantum effects become prominent
\citep{abbott2009observation,whittle2021approaching}. 

Although a purely classical treatment of the gravitational field still
suffices to explain all of the current experimental data it is nonetheless
interesting to ask what would be the quantum signature of gravitational
waves and several different theoretical approaches have been considered~\citep{calzetta1994noise,anastopoulos1996quantum,anastopoulos2013master,riedel2013evidence,blencowe2013effective,suzuki2015environmental,de2015decoherence,oniga2016quantum,oniga2017quantum,quinones2017quantum,vedral2020decoherence,xu2020toy}.
Recent theoretical works have also investigated the possibility of
detecting stochastic graviton noise in the context of gravitational
wave observatories~\citep{parikh2020noise,parikh2020quantum,parikh2020signatures}.

However, discerning between classical models of gravity from the quantum version
 will necessarily require testing coherent features of gravity
which cannot be mimicked by any classical noise source. Such a proposal
has been devised by considering the two nearby masses -- close enough
that they interact gravitationally but far enough apart that all other
channels of interaction are strongly suppressed -- which can entangle
only if the gravitational field exhibits bonafide quantum features~\citep{bose2017spin}~\footnote{The results of \cite{bose2017spin} were first reported in a conference talk in Bangalore \cite{ICTS}.}, see also~\cite{marletto2017gravitationally}.
The underlying mechanism for the quantum entanglement of masses (QGEM) has been analyzed within perturbative quantum gravity~\citep{Marshman:2019sne,bose2022mechanism, Vinckers:2023grv,Carney_2019,Carney:2021vvt}
and the framework of the Arnowitt--Desse--Meissner (ADM) approach~\citep{danielson2022gravitationally},
as well as in the path integral approach~\citep{christodoulou2023locally}, and for the massive graviton~\cite{Elahi:2023ozf}. Recent developments include an optomechanical proposal for testing the quantum light-bending interaction~\cite{Biswas:2022qto}, a quantum test of the weak equivalence principle~\cite{Bose:2022czr}, and test whether gravity acts as a quantum entity when measured~\cite{hanif2023testing}.

It is thus interesting to ask whether quantized gravitational waves
could also induce coherent effects in quantum systems, which would
be difficult to explain using a classical theory of gravity. 

In this work, we consider a quantum harmonic oscillator coupled to
quantized gravitational waves in the context of perturbative quantum
gravity. We first review the results of classical quadrupole radiation
emitted by a classical harmonic oscillator (Sec.~\ref{sec:Classical-quadrupole-radiation}).
We then obtain the matter-graviton coupling in the laboratory frame
of the quantum harmonic oscillator using Fermi normal coordinates~(Sec.~\ref{sec:construction}).
By tracing away the graviton, assumed to be in the vacuum state,
we obtain a simple master equation of the Lindblad form~\citep{gorini1976completely,lindblad1976generators}
for the quantum harmonic oscillator~(Sec.~\ref{sec:qft_model}).
The obtained dynamics have some important features. 
\begin{itemize}
    \item 
The total energy
of the quantum harmonic oscillator and of the emitted gravitons is
conserved~(Sec.~\ref{subsec:energyconservation}). 
\item The decay rate for number states $\vert n \rangle$ with $n\gg2$ 
is proportional to the square of its associated energy $E_n^2$ which can be seen as a consequence of Einstein's equivalence principle and of the quadrupole nature of gravitational waves~(Sec.~\ref{subsec:energy decay}).
\item For small occupation numbers, the classical and quantum predictions begin to differ, with the quantum harmonic oscillator retaining a steady-state coherence.
In particular, the quantum harmonic oscillator settles in a remnant coherent combination
of the ground and first excited states, which is a distinct quantum
signature of graviton emission~(Sec.~\ref{subsec:Coherence-protection-for-1}). 
\item For coherent states with large occupation numbers, we recover exactly the predictions
for a classical linear quandrupole. The obtained model can thus be seen as the quantum counterpart of the classical radiation theory~(Sec.~\ref{subsec:Recovering-classical-gravitation}).
\item The decoherence
rate scales with the cube of the trapping harmonic frequency, which
vanishes for a free particle as expected for a system without a mass
quadrupole. In particular, our model predicts that the center-of-mass of an isolated system never decoheres due to quantized gravitational waves~(Sec.~\ref{subsec:Coherence-protection-for}).
\end{itemize}
We conclude by discussing briefly the suitability of high frequency mechanical oscillator for distinguishing between classical and quantum gravitational waves (Sec.~\ref{summary}). In Appendix~\ref{sec:time-evolution-with} we provide the exact solution of the dynamics by formally mapping our problem to two-photon processes~\cite{simaan1978off,gilles1993two}, and in Appendix~\ref{sec:QGEM} we estimate the size of the effect for matter-wave interferometry focusing on the QGEM proposal.

\tableofcontents

\section{Classical quadrupole radiation\label{sec:Classical-quadrupole-radiation}}

\begin{figure}
\includegraphics{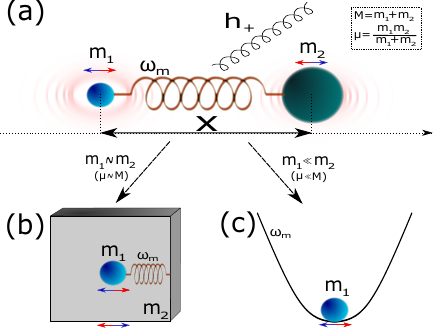} \caption{Graphical illustration of linear quadrupole radiation. (a) The linear
quadrupole is generated by the relative motion of two masses, $m_{1}$
and $m_{2}$, which are coupled with coupling $\omega_{\text{m}}$.
The center-of-mass motion (mass $M$) is unperturbed by the emission
of gravitational waves (which are of type ``+''), while the relative
motion (mass $\mu$) slowly decays as its energy is converted to gravitational
waves and radiated away. The two-particle problem can be always mapped
to the problem of a harmonically trapped reduced mass $\mu=\frac{m_{1}m_{2}}{m_{1}+m_{2}}$
with coupling $\sim\mu x^{2}$, where $x$ is relative distance between
the two masses. (b) Case $m_{1}\sim m_{2}$. A physical realization
consists of a trapped particle (i.e., mass $m_{1}$) inside a box
containing the apparatus to generate a harmonic trap (i.e., mass $m_{2}$).
The recoil of the system/box is on equal footing as $m_{1}\sim m_{2}$.
(c) Case $m_{1}\ll m_{2}$. The problem reduces to the motion of the
lighter mass $m_{1}$ in a harmonic trap with frequency $\omega_{\text{m}}$.
The recoil of the heavier mass $m_{2}$ is negligible with respect
to the recoil of the lighter mass $m_{1}$ (which can be approximately
identified with the reduced mass $\mu$) .}
\label{fig:quadrupole}
\end{figure}

We begin by briefly summarizing the main features of classical gravitational
radiation. We recall that gravitational radiation is sourced by a
time-dependent mass quadrupole. In this work, we are primarily interested
in the motion along one spatial axis where the simplest mass quadrupole
is given by a coupled two-particle system~\citep{de1986introduction}
(see Fig.~\ref{fig:quadrupole}). For concreteness we consider two
masses, $m_{1}$ and $m_{2}$, coupled by a quadratic potential: 
\begin{equation}
H_{\text{two-particle}}=\frac{(p^{\text{(1)}})^2}{2m_{1}}+\frac{(p^{\text{(2)}})^2}{2m_{1}}+\frac{k}{2}(x^{\text{(1)}}-x^{\text{(2)}})^{2},\label{eq:trap-1}
\end{equation}
where $k$ is the spring constant, and $x^{(1)}$ ($p^{(1)}$) and
$x^{(2)}$ ($p^{(2)}$) denote the position (momenta) of particle
1 and 2, respectively. It is useful to introduce the center-of-mass
coordinates: 
\begin{alignat}{2}
x & \equiv x^{(1)}-x^{(2)}, & p & \equiv p^{(1)}+p^{(2)},\label{eq:xp}\\
x_{\text{cm}} & \equiv\frac{m_{1}x^{(1)}+m_{2}x^{(2)}}{m_{1}+m_{2}},\,\,\, & p_{\text{cm}} & \equiv\frac{m_{2}p^{(1)}-m_{1}p^{(2)}}{m_{1}+m_{2}},\label{eq:xpcm}
\end{alignat}
as well as the reduced and total mass 
\begin{alignat}{1}
\mu & =\frac{m_{1}m_{2}}{m_{1}+m_{2}},\\
M & =m_{1}+m_{2},
\end{alignat}
respectively. Using the center-of-mass quantities we find that the
potential in Eq.~\eqref{eq:trap-1} reduces to 
\begin{equation}
H_{\text{\text{two-particle}}}=\frac{p_{\text{cm}}^{2}}{2M}+\frac{p^{2}}{2\mu}+\frac{\mu\omega_{\text{m}}^{2}}{2}x^{2},\label{eq:harmonic-2}
\end{equation}
where we have defined the harmonic frequency $\omega_{\text{m}}^{2}\equiv k/\mu$.

We make two well-known observations. On one hand, we note that the
center-of-mass remains uncoupled and is thus following a completely
free motion (i.e., in the general relativistic language the center-of-mass
position $x_{\text{cm}}$ follows a geodesic). On the other hand,
the relative motion is subject to a quadratic potential which can
give rise to a time-dependent linear quadrupole moment (and hence
can act as a source of gravitational radiation). In particular, we
consider the following relative motion (in the first instance neglecting
energy dissipation mechanisms): 
\begin{equation}
x=l\text{cos\ensuremath{(\omega_{m}t)}},\label{eq:x-1}
\end{equation}
where $l$ is the amplitude of oscillation of the relative motion.
The corresponding quadrupole moment tensor is given by 
\begin{equation}
D_{ij}=\int\rho(\bm{x}';\bm{x})\left(3x_{i}^{\prime}x_{j}^{\prime}-r^{\prime2}\delta_{ij}\right)d\bm{x}^{\prime},\label{eq:dij}
\end{equation}
where $\rho(\bm{x})$ is the mass density, $\bm{x}=(x_{1},x_{2},x_{3})$,
$i,j$ denote the spatial components, and $r^{2}=\sum_{i=1}^{3}x_{i}^{2}$.
Inserting in Eq.~\eqref{eq:dij} the mass density
\begin{equation}
\rho(\bm{x})=\mu\delta(x_{1}^{\prime}-x)\delta(x_{2}^{\prime})\delta(x_{3}^{\prime}),
\end{equation}
where $x$ is given in Eq.~\eqref{eq:x-1}, we readily find the following
non-vanishing elements: 
\begin{alignat}{1}
D_{11} & =D_{22}=-\frac{1}{2}D_{33}=-\mu l^{2}\text{cos\ensuremath{^{2}(\omega_{m}t)}}.
\end{alignat}
In particular, the linear quadrupole moment gives rise to gravitational
radiation of type ``+''. The average energy carried away by gravitational
waves is given by~\citep{de1986introduction,maggiore2008gravitational}:
\begin{equation}
\dot{E}=-\frac{16GI^{2}\omega_{\text{m}}^{6}}{15c^{2}},\label{eq:edecay}
\end{equation}
where we have introduced the moment of inertia $I=ml^{2}$. 
Thus as the two-particle system is oscillating it will slowly lose
energy -- the amplitude of the relative motion, $l$, will decay,
while the center-of-mass motion will remain completely unaffected.
Any quantum model of quantized gravitational waves should recover
the behaviour of classical quadrupole radiation when the state of
the harmonic oscillator can be modeled as approximately classical
(i.e., with a coherent state with a large occupation number).

\section{Linearised quantum gravity \label{sec:construction}}

In this section, our working hypothesis is linearised quantum gravity,
for a review see~\citep{Burgess:2003jk}. We further assume matter
to be non-relativistic, i.e. slowly moving. We find the dominant interaction
between matter and graviton in the Fermi normal coordinates (FNC)
which does not have any remnant gauge freedom~\citep{poisson2011motion,misner1973gravitation}
and is commonly used to describe laboratory experiments~\citep{will2006confrontation}.

\subsection{Fermi normal coordinates}

For simplicity, we will assume that the relevant motion of the particle
is along the $x$-axis and consider the FNC coordinates, $x^{\mu}=(-ct,x,y,z)$,
of an ideal observer following a geodesic trajectory (the situation
of an observer following a generic time-like curve can be analyzed
in a similar fashion). We start
from the general relativistic point-particle Lagrangian: 
\begin{equation}
L=-mc^{2}\sqrt{-g_{\mu\nu}\dot{x}^{\mu}\dot{x}^{\nu}},\label{eq:lagrangian}
\end{equation}
where $m$ is the mass of the system\footnote{Here $m$ denotes a generic mass. In the next Sec.~\ref{sec:twomass} we will have two such masses, $m_1$ and $m_2$ (see also Fig.~\ref{fig:quadrupole} for an illustration).}, $c$ is the speed of light,
and $g_{\mu\nu}$ is the metric expressed in FNC coordinates. In particular,
we write the metric as $g_{\mu\nu}=\eta_{\mu\nu}+h_{\mu\nu},$ $\eta_{\mu\nu}$
is the Minkowski metric, and $h_{\mu\nu}$ the spacetime curvature
perturbation near the geodesic up to order $\mathcal{O}(x^{2})$~\citep{visser2018post}.
Assuming that matter is moving slowly, the dominant contribution to
the dynamics will be given by~\citep{rakhmanov2014fermi}: 
\begin{equation}
g_{00}=-(1+\frac{1}{2c^{2}}\frac{\partial h_{11}^{2}}{\partial t^{2}}\vert_{t=0}x^{2}),\label{eq:gtt}
\end{equation}
where $\ddot{h}_{11}=2c^{2}R_{0101}$ is the ``+'' component of
the gravitational waves usually discussed in the transverse-traceless
(TT) coordinates, and $R_{0101}$ is the Riemann tensor component (here $\ddot{h}_{11}\vert_{t=0}\equiv\ddot{h}_{11}(t,0)$
denotes the value evaluated on the reference FNC curve). 

For completeness, let us sketch how to derive Eq.~\eqref{eq:field}.
One can decompose the graviational field into plane waves $\sim e^{+ikx}$
and Taylor expand then up to order $\mathcal{O}(x^{2})$:
\begin{equation}
h_{11}(t,x)\sim h(t,0)+\frac{\partial h(t,x)}{\partial x}\vert_{x=0}ikx-\frac{1}{2}\frac{\partial^{2}h(t,x)}{\partial x^{2}}\vert_{x=0}k^{2}x^{2},\label{eq:field}
\end{equation}
where $k=\omega_{k}/c$ and $\omega_{k}$ is the angular frequency
of the gravitational field mode. The first term on the RHS of Eq.~\eqref{eq:field}
is a constant and can be omitted, the second term $\sim kx$ vanishes
in FNC coordinates (i.e., linear potentials $\sim x$ vanish by choosing
an inertial reference frame), while the last term can be rewritten
as 

\begin{equation}
h_{11}(t,x)\sim-\frac{1}{2c^{2}}\frac{\partial^{2}h(t,x)}{\partial t^{2}}\vert_{t=0}x^{2},
\end{equation}
where we have used $\omega_{k}=kc$. For a rigorous derivation we
refer the reader to~\citep{rakhmanov2014fermi}.

Using Eqs.~(\ref{eq:lagrangian}) and (\ref{eq:gtt}) we then readily
find the interaction Lagrangian between graviton and matter degrees
of freedom. Finally, using $H_{\text{int}}=-L_{\text{int}}$ we find the interaction Hamiltonian
\begin{equation}
H_{\text{int}}=\frac{m}{4}\frac{\partial h_{11}^{2}}{\partial t^{2}}\vert_{t=0}x^{2}.\label{eq:intlagrangian}
\end{equation}
The quadratic coupling in Eq.~\eqref{eq:intlagrangian} has been
derived by assuming $k\Delta x\ll1$, where $\Delta x$ is the 
charachteristic size of the system (implicitly assumed in the expansion
in Eq.~\eqref{eq:field}). The latter condition can be rewritten
also as $\omega_{k}\Delta x/c\ll1$, where we have used $\omega_{k}=kc$ 
(i.e., we assume non-relativistic matter, with the trapped system moving much
more slowly than the speed of light).

\subsection{Harmonic oscillator coupled to gravitational waves}\label{sec:twomass}

From Eq.~\eqref{eq:intlagrangian} we find that for the two-particle system we have the following interaction Hamiltonian
\begin{equation}
H_{\text{int}}=\frac{1}{4}\ddot{h}_{11}\left[m_{1}(x^{\text{(1)}})^{2}+m_{2}(x^{\text{(2)}})^{2}\right],\label{eq:intlagrangian2}
\end{equation}
where $x^{\text{(1)}}$($x^{\text{(2)}}$) is the position of particle
1 (particle 2). We transform Eq.~\eqref{eq:intlagrangian2} using
the center-of-mass coordinates introduced in Eqs.~\eqref{eq:xp}
and \eqref{eq:xpcm} to find: 
\begin{equation}
H_{\text{int}}=\frac{1}{4}\ddot{h}_{11}\left[Mx_{\text{cm}}^{2}+\mu x^{2}\right]),\label{eq:intlagrangian3}
\end{equation}
where $M$ ($\mu$) is the total (reduced) mass.

From Eq.~\eqref{eq:intlagrangian3} it would thus appear that the
center-of-mass will be more strongly affected by the coupling to the
gravitational field than the relative motion (as we always have $M>\mu$
) -- we will show that this is \emph{not} the case. In particular,
we will find that only the relative degree of freedom can give rise
to graviton emission, whilst leaving undisturbed the center-of-mass
motion -- in complete analogy to classical quadrupole radiation.
Indeed, the quadrupole is linked to the harmonic trap term $\sim\omega_{\text{m}}^{2}x^{2}$ in Eq.~\eqref{eq:harmonic-2}, while the center-of-mass motion is unconstrained and can thus only
follow a geodesic (i.e., it is in free-fall) and as such cannot radiate.

\subsection{Matter-graviton Hamiltonian \label{subsec:Hamiltonian}}

We now obtain the leading order coupling between gravitons and harmonically
trapped quantum matter. We consider the gravitational field expanded
in plane waves~\citep{Weinberg:1972kfs,oniga2016quantum}: 
\begin{equation}
\hat{h}_{ij}(t,\bm{x})=\int d\bm{k}\sqrt{\frac{G\hbar}{\pi^{2}c^{2}\omega_{k}}}\hat{g}_{\bm{k},\lambda}\mathtt{e}_{ij}^{\lambda}(\bm{n})e^{-i(\omega_{k}t-\bm{k}\cdot\bm{x})}+\text{H.c.},\label{eq:gravfield}
\end{equation}
where $G$ is the Newton's constant, $\omega_{k}=kc$, $k=\|\bm{k}\|$,
$\bm{n}=\bm{k}/\|\bm{k}\|$, and $\hat{g}_{\bm{k},\lambda}$ is the
annihilation operator. In Eq.~\eqref{eq:gravfield} we also implicitly
assume the summation over the polarizations, $\sum_{\lambda}$, where
$\mathtt{e}_{jk}^{\lambda}$ denote the basis tensors for the two
polarizations, $\lambda=1,2$. The basis tensors satisfy the completeness
relation: 
\begin{equation}
\sum_{\lambda}\mathtt{e}_{ij}^{\lambda}(\bm{n})\mathtt{e}_{kl}^{\lambda}(\bm{n})=P_{ik}P_{jl}+P_{il}P_{jk}-P_{ij}P_{kl},\label{eq:completeness}
\end{equation}
where $P_{ij}\equiv P_{ij}(\bm{n})=\delta_{ij}-\bm{n}_{i}\bm{n}_{j}$.
From Eq.~\eqref{eq:intlagrangian} and \eqref{eq:gravfield} we however
see that only $\mathtt{e}_{11}^{\lambda}(\bm{n})$ is relevant for
the matter-wave system. For later convenience we write the integral:
\begin{equation}
\int d\bm{n}\,P_{11}(\bm{n})P_{11}(\bm{n})=\frac{32\pi}{15}.\label{eq:angles}
\end{equation}
As we will see this latter expression quantifies the average effect
(on the $x$-axis motional state of the harmonic oscillator) of the
gravitons emitted in arbitrary directions.

We can readily write also the corresponding kinetic term for the massless
graviton field: 
\begin{equation}
H_{\text{grav}}=\int d\bm{k}\,\hbar\omega_{k}g_{\bm{k},\lambda}^{\dagger}\hat{g}_{\bm{k},\lambda}.\label{eq:Hgrav}
\end{equation}
In addition, we assume that the matter degree of freedom is harmonically
trapped and described by the following Hamiltonian\footnote{We will first focus on the coupling of the relative motion, $x$, to the gravitational field. The results for the center-of-mass, $x_{\text{cm}}$,
can then be obtained by taking the limit $\omega_{\text{m}}\rightarrow0$.
In this limit the equations for the center-of-mass and relative motion become
of the same form as can be concluded from the Hamiltonians in Eqs.~\eqref{eq:harmonic-2} and \eqref{eq:intlagrangian3}.}

\begin{equation}
H_{\text{m}}=\frac{\hat{p}^{2}}{2m}+\frac{\mu\omega_{\text{m}}^{2}}{2}\hat{x}^{2},\label{eq:Hmfirst}
\end{equation}
where $\omega_{\text{m}}$ is the harmonic frequency. As we will see
it is convenient to introduce the (adimensional) amplitude quadrature,
\begin{equation}
\hat{X}=\hat{b}+\hat{b}^{\dagger},\,\label{eq:X}
\end{equation}
which is related to the position observable as $\hat{x}=\delta_{\text{zpf}}\hat{X}$,
and the matter-zero-point-fluctuations are given by 
\begin{equation}
\delta_{\text{zpf}}=\sqrt{\frac{\hbar}{2\mu\omega_{\text{m}}}}.\label{zero-point}
\end{equation}
Furthermore, we introduce the (adimensional) phase quadrature

\begin{equation}
\hat{P}=i(\hat{b}^{\dagger}-\hat{b})\,\label{eq:P}
\end{equation}
which is related to the physical momentum observable as $\hat{p}=\sqrt{\frac{\hbar \mu\omega_{\text{m}}}{2}}\hat{P}$.
In particular, we can rewrite Eq.~\eqref{eq:Hmfirst} in the standard
notation:

\begin{equation}
\hat{H}_{\text{m}}=\frac{\hbar\omega_{\text{m}}}{4}(\hat{X}^{2}+\hat{P}^{2})=\hbar\omega_{\text{m}}\hat{b}^{\dagger}\hat{b}.\label{eq:Hm}
\end{equation}

The interaction Hamiltonian is given in Eq.~(\ref{eq:intlagrangian}), and using Eq.~(\ref{eq:gravfield})
we find: 
\begin{alignat}{1}
H_{\text{int}} & =\sum_{\lambda}\int d\bm{k}\,\mathcal{G}_{\bm{k}}^{\lambda}\hat{g}_{\bm{k},\lambda}\hat{X}^{2}\text{+H.c.},\label{eq:Hint}
\end{alignat}
where the coupling is given by 
\begin{equation}
\mathcal{G}_{\bm{k}}^{\lambda}=\sqrt{\frac{G\hbar^{3}\omega_{k}^{3}}{64\pi{}^{2}c^{2}\omega_{\text{m}}^{2}}}\mathtt{e}_{11}^{\lambda}(\bm{n}).\label{eq:coupling}
\end{equation}
Importantly, we note that the coupling $\mathcal{G}_{\bm{k}}^{\lambda}$
in Eq.~\eqref{eq:coupling} does not depend on the mass of the matter
system, but only on graviton and matter-wave frequencies, $\omega_{\bm{k}}$
and $\omega_{m}$, respectively -- as such, the effect on the matter
system in the mesoscopic is precisely the same as, say, on atomic
systems. Of course, for a given value of the position amplitude $\hat{X}$,
since $\hat{x}\sim\hat{X}/\sqrt{\mu}$, the lighter system will have
a larger physical position in comparison to the heavier one. Similarly,
if we would express the harmonic frequency, $\omega_{\text{m}},$
in terms of the spring contant, $K=\omega_{\text{m}}^{2}/\mu$, we would
again find a dependency of the coupling on the mass of the system
$\mu$.

In summary, the total Hamiltonian is now given by: 
\begin{equation}
\hat{H}_{\text{tot}}=\hat{H}_{\text{grav}}+\hat{H}_{\text{m}}+\hat{H}_{\text{int}},
\end{equation}
where $\hat{H}_{\text{grav}}$, $\hat{H}_{\text{m}}$, and $\hat{H}_{\text{int}}$
are given in Eqs.~(\ref{eq:Hgrav}), (\ref{eq:Hm}), and (\ref{eq:Hint}).

\section{Decoherence due to graviton bath -- QFT model\label{sec:qft_model}}

In this section, we will obtain the dynamics of the matter system when
coupled to the quantum field model of the gravitational field (see
the previous Sec.~\ref{sec:construction}). We will refer to the
developed model as the QFT model (the gravitational field will be
considered from the QFT point of view, while the matter system will
be modeled in the first quantization). 
We will assume that the graviton field is in the ground state, i.e.
without any excitations (i.e., an initially empty bath): 
\begin{alignat}{1}
\langle\hat{g}_{\bm{k},\lambda}^{\dagger}\hat{g}_{\bm{k}',\lambda'}\rangle & =0,\label{eq:gg}\\
\langle\hat{g}_{\bm{k},\lambda}\hat{g}_{\bm{k}',\lambda'}^{\dagger}\rangle & =\delta^{(3)}(\bm{k}-\bm{k}')\delta_{\lambda,\lambda'},\label{eq:ggd}
\end{alignat}
and $\langle\hat{g}_{\bm{k},\lambda}^{\dagger}\hat{g}_{\bm{k}',\lambda'}^{\dagger}\rangle=\langle\hat{g}_{\bm{k},\lambda}\hat{g}_{\bm{k}',\lambda'}\rangle=0$.

We construct the quantum master equation for the matter system
-- by tracing out the gravitational field -- closely following the
generic derivation from~\citep{breuer2002theory} (see Chapter 3.3).
We denote the total statistical operator of the problem as $\hat{\rho}^{\text{(tot)}}$
(the matter-wave system and the gravitational field), and by $\hat{\rho}$
($\hat{\rho}^{(\text{g})}$) the reduced statistical operator for
the matter-wave system (the gravitational field), obtained by tracing
away the gravitational field (the matter-wave system). The von-Neumann
equation can be expressed in the interaction picture as: 
\begin{equation}
\frac{d}{dt}\hat{\rho}_{t}^{\text{(tot)}}=-\frac{i}{\hbar}[H_{t}^{\text{(int)}},\hat{\rho}_{t}^{\text{(tot)}}],\label{eq:vonNeumann}
\end{equation}
where 
\begin{alignat}{1}
\hat{H}_{t}^{\text{(int)}}= & \sum_{\lambda}\int d\bm{k}\,\mathcal{G}_{\bm{k}}^{\lambda}\hat{g}_{\bm{k},\lambda}e^{-i\omega_{\bm{k}}t}\,\hat{X}_{t}^{2}+\text{H.c.}\label{eq:Hintint}
\end{alignat}
is the interaction Hamiltonian from Eq.~\eqref{eq:Hint} transformed
into the interaction picture. The amplitude quadrature (the adimensional
position observable) in the interaction picture is given by: 
\begin{equation}
\hat{X}_{t}=\hat{b}e^{-i\omega_{\text{m}}t}+\hat{b}^{\dagger}e^{i\omega_{\text{m}}t}.\label{eq:amplitude}
\end{equation}
The dynamics in Eq.~\eqref{eq:vonNeumann} can be formally solved:
\begin{equation}
\hat{\rho}_{t}^{\text{(tot)}}=\hat{\rho}_{0}^{\text{(tot)}}-\frac{i}{\hbar}\int_{0}^{t}ds\,[\hat{H}_{s}^{\text{(int)}},\hat{\rho}_{s}^{\text{(tot)}}].\label{eq:formalsolution}
\end{equation}
By then inserting Eq.~\eqref{eq:formalsolution} into Eq.~\eqref{eq:vonNeumann},
and tracing over the bath (the gravitational field), we obtain: 
\begin{equation}
\frac{d}{dt}\hat{\rho}_{t}=-\frac{1}{\hbar^{2}}\int_{0}^{t}ds\,\text{tr}_{\text{g}}[\hat{H}_{t}^{\text{(int)}},[\hat{H}_{s}^{\text{(int)}},\hat{\rho}_{s}^{\text{(tot)}}]],\label{eq:secondorder}
\end{equation}
where the first-order term, $\text{tr}_{\text{g}}[H_{t}^{\text{(int)}},\hat{\rho}_{0}^{\text{(tot)}}]$,
vanished as $\langle\hat{g}_{\bm{k},\lambda}\rangle=\langle\hat{g}_{\bm{k},\lambda}^{\dagger}\rangle=0$.
On the other hand, the second-order term on the right-hand side of
Eq.~\eqref{eq:secondorder} is non-zero -- as it depends on the
value of the vacuum fluctuations, $\langle\hat{g}_{\bm{k},\lambda}\hat{g}_{\bm{k}',\lambda'}^{\dagger}\rangle$,
defined in Eq.~\eqref{eq:ggd}. Eq.~\eqref{eq:secondorder} is,
however, still a formal (exact) relation, containing the net effect
of all Feynman diagrams with any number of vertices. We will now discuss
the approximations that will lead to the more familiar Lindblad form
of the quantum master equation -- describing the effect of the dominant
tree-level Feynman diagram contributions -- exploiting the weakness
of the coupling $\sim\sqrt{G}$. 

We first impose the Born approximation, $\hat{\rho}_{s}^{\text{(tot)}}\approx\hat{\rho}_{s}\otimes\hat{\rho}^{(\text{g})}$,
on the right hand-side of Eq.~\eqref{eq:secondorder}. Importantly,
the Born approximation precludes from the analysis any entanglement
between the matter-wave system, and the gravitational field as we
are explicitly assuming a factorizable state. Furthermore, the state
of the graviton bath, $\hat{\rho}^{(\text{g})}\equiv\hat{\rho}_{0}^{(\text{g})}$,
is always the same as far as the system is concerned -- here we are
assuming that the interaction between the system and the gravitational
field is weak, with negligible effect on the latter.

We next want to make Eq.~\eqref{eq:secondorder} local in time (i.e.,
such that the dynamics will depend only on the state $\hat{\rho}_{t}$
at time $t$, but not on the state at earlier times) and independent
of the choice of the initial time. To this end, we first formally solve
the von Neumann equation to connect the state at time $s$ with the
state at time $t$ to find:

\begin{equation}
\hat{\rho}_{s}^{\text{(tot)}}=\hat{\rho}_{t}^{\text{(tot)}}+\frac{i}{\hbar}\int_{s}^{t}ds'\,[\hat{H}_{s'}^{\text{(int)}},\hat{\rho}_{s'}^{\text{(tot)}}],\label{eq:formalsolution2}
\end{equation}
i.e., similarly, as we have done in Eq.~\eqref{eq:formalsolution}.
We then insert Eq.~\eqref{eq:formalsolution2} into Eq.~\eqref{eq:secondorder}
to find:

\begin{alignat}{1}
\frac{d}{dt} & \hat{\rho}_{t}=-\frac{1}{\hbar^{2}}\int_{0}^{t}ds\underbrace{\text{tr}_{\text{g}}[\hat{H}_{t}^{\text{(int)}},[\hat{H}_{s}^{\text{(int)}},\hat{\rho}_{t}^{\text{(tot)}}]]}_{\sim G\,\text{terms}}\nonumber \\
 & -\frac{i}{\hbar^{3}}\int_{0}^{t}ds\int_{s}^{t}ds'\underbrace{\text{tr}_{\text{g}}[\hat{H}_{t}^{\text{(int)}},[\hat{H}_{s}^{\text{(int)}},[\hat{H}_{s'}^{\text{(int)}},\hat{\rho}_{s'}^{\text{(tot)}}]]]}_{\sim G^{3/2}\,\text{and higher order terms}},\label{eq:twolines}
\end{alignat}
where we have explicitly separated the dominant terms $\sim G$ in
the first line from the higher order terms in the second line (each
Hamiltonian operator introduces a vertex $\sim\sqrt{G}$). In the
following we will truncate the dynamics at the dominant order $\sim G$
which is justified by the weakness of the graviton-matter coupling
$\sim\sqrt{G}$ (see Eqs.~\eqref{eq:coupling} and \eqref{eq:Hintint}).
In this way, we find an equation that is local in time (i.e., it depends
only on the state $\hat{\rho}_{t}$ at time $t$, but not on the state
at earlier times). In addition, we change the integration variable
$s\rightarrow t-s$ and extend the integration limit to infinity,
i.e., $\int_{0}^{t}\rightarrow\int_{0}^{\infty}$, to find:

\begin{equation}
\frac{d}{dt}\hat{\rho}_{t}=-\frac{1}{\hbar^{2}}\int_{0}^{\infty}ds\,\text{tr}_{\text{g}}[H_{t}^{\text{(int)}},[H_{t-s}^{\text{(int)}},\hat{\rho}_{t}^{\text{(tot)}}]].\label{eq:integrationlimit}
\end{equation}
Extending the integration limit to infinity is allowed if the integrand
decays sufficiently fast for values of $s$ different from $t$ --
such an assumption is valid when the decay time of the graviton bath
correlation function is much faster than the time scale over which
the state of the system changes appreciably. This makes the dynamics
in Eq.~\eqref{eq:integrationlimit} independent of the choice for
the initial time (compare with Eq.~\eqref{eq:secondorder}), which
is a sensible requirement for non-relativistic matter coupled to
the gravitational field. The steps in Eqs.~\eqref{eq:twolines} and
\eqref{eq:integrationlimit} is equivalent to taking the Markov approximation
commonly performed in analogous electromagnetic calculations (see
Refs. \citep{louisell1970quantum,breuer2002theory,gardiner2004quantum,schlosshauer2007decoherence}
for more details).

In summary, applying the approximations from the previous two paragraphs
to Eq.~\eqref{eq:secondorder}, we find the following Markovian master
equation: 
\begin{equation}
\frac{d}{dt}\hat{\rho}_{t}=-\frac{1}{\hbar^{2}}\int_{0}^{\infty}ds\,\text{tr}_{\text{g}}[H_{t}^{\text{(int)}},[H_{t-s}^{\text{(int)}},\hat{\rho}_{t}\otimes\hat{\rho}^{(\text{g})}]].\label{eq:secondorder2}
\end{equation}
We then proceed by inserting the interaction Hamiltonian from Eq.~\eqref{eq:Hintint}
into Eq.~\eqref{eq:secondorder2} to eventually find 
\begin{alignat}{1}
\frac{d}{dt}\hat{\rho}_{t}= & -\frac{1}{\hbar^{2}}\int_{0}^{\infty}ds\int d\bm{k}\int d\bm{k}'\,\nonumber \\
 & \times\langle\hat{g}_{\bm{k},\lambda}\hat{g}_{\bm{k}',\lambda'}^{\dagger}\rangle\,\mathcal{G}_{\bm{k}}^{\lambda}\mathcal{G}_{\bm{k}'}^{\lambda'}\,e^{-i(\omega_{k}-\omega_{k'})t}\nonumber \\
 & \times\Bigl\{ e^{-i\omega_{k'}s}\hat{X}_{t}^{2}\hat{X}_{t-s}^{2}\hat{\rho}_{t}-e^{i\omega_{k'}s}\hat{X}_{t}^{2}\hat{\rho}_{t}\hat{X}_{t-s}^{2}\nonumber \\
 & \qquad e^{i\omega_{k}s}\hat{\rho}_{t}\hat{X}_{t-s}^{2}\hat{X}_{t}^{2}-e^{-i\omega_{k}s}\hat{X}_{t-s}^{2}\hat{\rho}_{t}\hat{X}_{t}^{2}\Bigr\},
\end{alignat}
where we have already used the fact that there are no excitations
of the gravitational field, $\langle\hat{g}_{\bm{k},\lambda}^{\dagger}\hat{g}_{\bm{k}',\lambda'}\rangle=0$
(see Eq.~\eqref{eq:gg}).

We now insert the non-zero value for the vacuum fluctuations, $\langle\hat{g}_{\bm{k},\lambda}\hat{g}_{\bm{k}',\lambda'}^{\dagger}\rangle\sim\delta(\bm{k}-\bm{k}')\delta_{\lambda,\lambda'}$
(see Eq.~\eqref{eq:ggd}): 
\begin{alignat}{1}
\frac{d}{dt}\hat{\rho}_{t}= & -\frac{1}{\hbar^{2}}\int_{0}^{\infty}ds\int d\bm{k}\,\sum_{\lambda}(\mathcal{G}_{\bm{k}}^{\lambda})^{2}\nonumber \\
 & \times\Bigl\{ e^{-i\omega_{k}s}\hat{X}_{t}^{2}\hat{X}_{t-s}^{2}\hat{\rho}_{t}-e^{i\omega_{k}s}\hat{X}_{t}^{2}\hat{\rho}_{t}\hat{X}_{t-s}^{2}\nonumber \\
 & \qquad e^{i\omega_{k}s}\hat{\rho}_{t}\hat{X}_{t-s}^{2}\hat{X}_{t}^{2}-e^{-i\omega_{k}s}\hat{X}_{t-s}^{2}\hat{\rho}_{t}\hat{X}_{t}^{2}\Bigr\},
\end{alignat}
and inserting the expression for the coupling $\mathcal{G}_{\bm{k}}^{\lambda}$
from Eq.~\eqref{eq:coupling}, to obtain: 
\begin{alignat}{1}
\frac{d}{dt}\hat{\rho}_{t}= & -\int_{0}^{\infty}ds\int d\bm{k}\,\frac{G\hbar\omega_{k}^{3}}{64\pi{}^{2}c^{2}\omega_{m}^{2}}\sum_{\lambda}\mathtt{e}_{11}^{\lambda}(\bm{k})\mathtt{e}_{11}^{\lambda}(\bm{k})\nonumber \\
 & \times\Bigl\{ e^{-i\omega_{k}s}\hat{X}_{t}^{2}\hat{X}_{t-s}^{2}\hat{\rho}_{t}-e^{i\omega_{k}s}\hat{X}_{t}^{2}\hat{\rho}_{t}\hat{X}_{t-s}^{2}\nonumber \\
 & \qquad e^{i\omega_{k}s}\hat{\rho}_{t}\hat{X}_{t-s}^{2}\hat{X}_{t}^{2}-e^{-i\omega_{k}s}\hat{X}_{t-s}^{2}\hat{\rho}_{t}\hat{X}_{t}^{2}\Bigr\}. \label{eq:megamtt}
\end{alignat}
The summation can be evaluated using the completeness relation from
Eq.~\eqref{eq:completeness} and the relation in Eq.~\eqref{eq:angles}
-- we then integrate over the solid angle by first expressing the
integration measure as $d\bm{k}=k^{2}dkd\bm{n}=\frac{\omega_{k}^{2}}{c^{3}}d\omega_{k}d\bm{n}$,
where $k=\|\bm{k}\|$ and $\bm{n}=\bm{k}/\|\bm{k}\|$. From Eq.~\eqref{eq:megamtt} we thus find:
\begin{alignat}{1}
\frac{d}{dt}\hat{\rho}_{t}= & -\int_{0}^{\infty}ds\int_{0}^{\infty}d\omega_{k}\,\frac{G\hbar\omega_{k}^{5}}{30\pi c^{5}\omega_{\text{m}}^{2}}.\nonumber \\
 & \times\Bigl\{ e^{-i\omega_{k}s}\hat{X}_{t}^{2}\hat{X}_{t-s}^{2}\hat{\rho}_{t}-e^{i\omega_{k}s}\hat{X}_{t}^{2}\hat{\rho}_{t}\hat{X}_{t-s}^{2}\nonumber \\
 & \qquad e^{i\omega_{k}s}\hat{\rho}_{t}\hat{X}_{t-s}^{2}\hat{X}_{t}^{2}-e^{-i\omega_{k}s}\hat{X}_{t-s}^{2}\hat{\rho}_{t}\hat{X}_{t}^{2}\Bigr\}.
\end{alignat}
We now finally insert the position amplitude observable from Eq.~\eqref{eq:amplitude},
and apply the rotating wave approximation, i.e. we keep terms with
equal number of $\hat{b}$ and $\hat{b}^{\dagger}$, and neglect the
other fast rotating terms which typically give only a small correction,
see~\citep{agarwal2012quantum}: 
\begin{alignat}{1}
\frac{d}{dt}\hat{\rho}_{t}= & -\int_{0}^{\infty}d\omega_{k}\,\frac{G\hbar\omega_{k}^{5}}{30\pi c^{5}\omega_{\text{m}}^{2}}\int_{0}^{\infty}ds\,\nonumber \\
 & e^{-i(\omega_{k}-2\omega_{\text{m}})s}\left(\hat{b}^{\dagger2}\hat{b}^{2}\hat{\rho}_{t}-\hat{b}^{2}\hat{\rho}_{t}\hat{b}^{\dagger2}\right)\nonumber \\
 & +e^{i(\omega_{k}-2\omega_{\text{m}})s}\left(\hat{\rho}_{t}\hat{b}^{\dagger2}\hat{b}^{2}-\hat{b}^{2}\hat{\rho}_{t}\hat{b}^{\dagger2}\right),\label{eq:46}
\end{alignat}
where we have kept only the non-zero contribution, $\sim\delta(\omega_{s}-\omega_{k})$,
while we have omitted the other contributions. Specifically, the terms
$\sim\delta(\omega_{s}+\omega_{k})$ are zero as the graviton cannot
have negative frequency, while the terms $\sim\omega_{k}^{5}\delta(\omega_{k})$
vanish. We then finally integrate over all possible out-going graviton
frequencies, $\omega_{k}$, using the fact that
\begin{equation}\label{eq:conserv}
    \int_{0}^{\infty}ds\,e^{-i(\omega_{k}-2\omega_{\text{m}})s}=\pi\delta(\omega_{k}-2\omega_{\text{m}}),
\end{equation}
i.e., the QFT and open quantum system formalism leads us the energy conservation (see Sec.~\ref{subsec:energyconservation} for a discussion). Eventually, we find a simple Lindblad equation (in Schr\"{o}dinger picture):
\begin{equation}
\frac{d}{dt}\hat{\rho}_{t}=\gamma_{\text{grav}}\left(\hat{b}^{2}\hat{\rho}_{t}\hat{b}^{\dagger2}-\frac{1}{2}\{\hat{b}^{\dagger2}\hat{b}^{2},\hat{\rho}_{t}\}\right),\label{eq:lindblad}
\end{equation}
where $\{\,\cdot\,,\,\cdot\,\}$ denotes the anti-commutator, and
the emission rate is given by
\begin{equation}
\gamma_{\text{grav}}=\frac{32}{15}t_{\text{Pl}}^{2}\omega_{\text{m}}^{3}.\label{eq:gammagrav}
\end{equation}
$\gamma_{\text{grav}}$ is parameter-free and depends on fundamental
constants of nature only through the Planck time, $t_{\text{Pl}}=(G\hbar/c^{5})^{1/2}\sim10^{-43}~\text{s}$, and is independent of the mass or any other intrinsic or extrinsic
property of the system apart from the frequency $\omega_{\text{m}}$. Of course, if one would express $\omega_{\text{m}}$ in terms of
the mechanical spring constant, $K=\omega_{\text{m}}^{2}/\mu$, then $\gamma_{\text{grav}}$ would depend on the mass of the system as
$\mu^{3/2}$.

The quantum master equation in Eq.~\eqref{eq:lindblad} is valid
for a wide range of particle masses, from the microscopic, e.g., optically trapped atoms, to the mesoscopic scale and beyond, e.g., the $10$ kg LIGO mirror. In particular, it can
be used to estimate the gravitational decoherence for any harmonically
trapped system\footnote{We recall that in deriving Eq.~\eqref{eq:lindblad} we have implicitly
performed the calculation for
long-wavelength gravitons (with wavelength $\lambda$ large compared
to the spatial delocalization $\Delta x$) as for typical experimental
frequencies $\omega_{\text{m}}$ and delocalizations $\Delta x$
we always have $\lambda=\frac{2\pi c}{\omega_{\text{m}}}\gg\Delta x$.
We leave the calculation for short-wavelength gravitons
for future research (we would need to consider higher-order FNC terms
in Eqs.~\eqref{eq:gtt}-\eqref{eq:intlagrangian}).}.

\section{Consequences of the QFT model~\label{sec:Consequences}}
In the previous Sec.~\ref{sec:qft_model} we have derived the dynamics for a harmonically trapped system coupled to an initially empty graviton bath. At the leading order $\mathcal{O}(G)$ and $\mathcal{O}(c^{-2})$   we have found the master equation in Eq.~\eqref{eq:lindblad} which describes the dynamics of the matter system as it emits gravitons. In Secs.~\ref{subsec:energyconservation} - \ref{subsec:Coherence-protection-for} we now make a series of key observations about the derived QFT model.

\subsection{Conservation of total energy~\label{subsec:energyconservation}}
We first highlight that the total energy of the system, formed by the matter and the gravitational field, is conserved in the derived QFT model. 
The starting point of the analysis was the interaction Hamiltonian in Eq.~(\ref{eq:Hint}), which is energy conserving (i.e., we have an associated energy conserving Feynman vertex diagram $\sim\sqrt{G}$).

To illustrate in more detail the energy balance let us consider energy eigenstates of the harmonic oscillator with the energy levels separated by multiples of $\hbar\omega_\text{m}$.
The matter system of initial
energy $\hbar \omega_{\text{i}}$ emits an \emph{on-shell} graviton of energy
$\hbar\omega_{k}=2 \hbar \omega_{\text{m}}$ resulting in the final matter energy $\hbar \omega_{\text{f}}=\hbar\omega_{\text{i}}-2\hbar\omega_{\text{m}}$. The graviton frequency $2\omega_{\text{m}}$ is a consequence of the \emph{quadratic}
position coupling of the matter system, $\sim\hat{X}^{2}\propto (\hat{b}+\hat{b}^\dagger)^2$, to
the gravitational field, $\hat{g}_{\bm{k},\lambda}$. However, we did not impose energy conservation
at any stage, but rather the energy balance arises directly from the matter-graviton coupling and the quantum
field theory analysis. In particular, see how the energy conserving condition $\delta(\omega_{k}-2\omega_{\text{m}})$
emerges in Eq.~\eqref{eq:conserv}.

The energy balance can thus be summarized as:
\begin{equation}
\underbrace{\hbar\,\omega_{\text{i}}-\,\hbar\,\omega_{\text{f}}}_{\text{matter}}=2\hbar\omega_{\text{m}}=\underbrace{\hbar\omega_{k}}_{\text{graviton}}.\label{eq:balance}
\end{equation}
Importantly, as $\omega_{\text{m}}>0$ the matter system loses energy,
i.e. $\omega_{\text{f}}<\omega_{\text{i}}$. The energy of the matter
subsystem is monotonously decreasing with the energy carried away
by the emitted gravitons, but the total energy of the matter-graviton
system remains conserved.

We can also understand why graviton emission is the only possible
process at order $\sim\sqrt{G}$  based on physical considerations. As the gravitational field is initially in the lowest energy state $\vert0\rangle$ it can only absorb energy from the matter-system
(i.e., graviton emission from the matter system) while all other
processes are forbidden by energy conservation.

\subsection{Decay of number states\label{subsec:energy decay}}
In the following we will be interested in the number states $\vert n \rangle$. For such states we can estimate the order of magnitude of the decoherence by computing the decay of the phonon number: 
\begin{equation}
\frac{d}{dt}\langle\hat{n}\rangle_{t}=\text{tr}\left[\hat{n}\frac{d}{dt}\hat{\rho}_{t}\right],
\end{equation}
where $\langle\,\cdot\,\rangle_{t}=\text{tr}[\,\cdot\,\rho_{t}]$,
and $\hat{n}=\hat{b}^{\dagger}\hat{b}$. Specifically, inserting the
QFT model from Eq.~\eqref{eq:lindblad}, and using the cyclic property
of the trace, we readily find: 
\begin{alignat}{1} \label{eq:aaa}
\frac{d}{dt}\langle\hat{n}\rangle_{t} & =\gamma_{\text{grav}}\langle\hat{b}^{\dagger2}\hat{b}^{\dagger}\hat{b}\hat{b}^{2}-\frac{1}{2}\{\hat{b}^{\dagger2}\hat{b}^{2},\hat{b}^{\dagger}\hat{b}\}\rangle_{t}.
\end{alignat}
From Eq.~\eqref{eq:aaa}, recalling the commutation relation $[\hat{b},\hat{b}^{\dagger}]=1$, we
then eventually find: 
\begin{equation}
\frac{d}{dt}\langle\hat{n}\rangle_{t}=-2\gamma_{\text{grav}}\langle\hat{b}^{\dagger2}\hat{b}^{2}\rangle_{t}.\label{eq:energydecay}
\end{equation}
Finally, using the identity $\hat{b}^{\dagger 2}\hat{b}^{2}=\hat{b}^{\dagger} \hat{b}\hat{b}^{\dagger}\hat{b} - \hat{b}^{\dagger} [\hat{b},\hat{b}^{\dagger}]\hat{b}$ (and again the  commutation relation) we then obtain from Eq.~\eqref{eq:energydecay}:
\begin{equation}
\frac{d}{dt}\langle\hat{n}\rangle_{t}=-2\gamma_{\text{grav}}\left(\langle \hat{n}^2\rangle_{t} -\langle \hat{n} \rangle_t\right).\label{eq:energydecay222}
\end{equation}
We first note that for the number state $\vert n \rangle$ with $n\geq2$ the right-hand side of Eq.~\eqref{eq:energydecay222} is negative (i.e., the number state decays). In particular, for  $n\gg 2$ we can neglect the term $\langle \hat{n} \rangle_t$ and we can rewrite Eq.~\eqref{eq:energydecay222} as 
\begin{equation}
\dot{n}\approx-\frac{64}{15}\omega_\text{m} \left( \frac{E_n}{E_\text{Pl}}\right)^2,\label{eq:energydecay-22}
\end{equation}
where we have used the definition of $\gamma_\text{grav}$ from Eq.~\eqref{eq:gammagrav}, the relation between Planck time and energy, $t_\text{Pl}=\hbar/E_\text{Pl}$, 
and the standard definition of the harmonic oscillator energy, $E_n=\hbar\omega_\text{m} n$. The quadratic dependence on the energy in Eq.~\eqref{eq:energydecay222} can be seen as a consequence of Einstein's equivalence principle and of the quadrupole nature of gravitational waves which was the starting point of our analysis (see discussion of Eqs.~\eqref{eq:lagrangian}-\eqref{eq:intlagrangian}).

We second note that for $\vert n \rangle$ with $n=0,1$ the decay rate in Eq.~\eqref{eq:energydecay222} is zero (i.e., the number states $\vert 1\rangle$ and $\vert 0\rangle$ do not decay). This is hinting at the idea of coherence protection from quantized gravitational waves, which we now discuss in Sec.~\ref{subsec:Coherence-protection-for-1}.

\subsection{Coherence protection for the relative-motion\label{subsec:Coherence-protection-for-1}}

Let us now discuss the consequences of the QFT model when we have
low occupation numbers. The master equation in Eq.~\eqref{eq:lindblad} is formally equivalent to the master equation appearing in the context of two-photon absorption problems~\cite{simaan1978off}. Exploiting this formal mapping we find that Eq.~\eqref{eq:lindblad} can be solved analytically with the steady-state given by:
\begin{alignat}{1}
\hat{\rho}_{\infty} & =\lambda_{0}\vert0\rangle\langle0\vert+\lambda_{1}\vert1\rangle\langle1\vert+\lambda_{c}\left[\vert0\rangle\langle1\vert+\vert1\rangle\langle0\vert\right],\label{eq:finalstate-1}
\end{alignat}
where $\lambda_{0}$,$\lambda_{1}$ and $\lambda_{c}$ depend on
the initial matter state $\hat{\rho}_{0},$ and $\vert\,\cdot\,\rangle$
denotes a number state. The coefficients $\lambda_j$ ($j=0,1,c$) are given for completeness in Appendix~\ref{sec:time-evolution-with} following~\cite{gilles1993two}. 

When discussing a coupling to an empty gravitational bath the naive
expectation would have been that the system will eventually decay to the ground
state emitting all of its energy -- in line with the energy decay
predicted by the classical gravitational radiation formula given in Eq.~\eqref{eq:edecay}. 
However, as we start approaching the ground state the nature
of the two-phonon process $\sim\hat{b}^{2}$ in Eq.~\eqref{eq:lindblad}
begins to modify the continuous classical picture. Indeed, applying
twice the annihilation operator $\hat{b}$ to the number state $\vert n\rangle$
induces the transition:
\begin{equation}
\vert n\rangle\rightarrow\vert n-2\rangle,\label{eq:twophonon}
\end{equation}
and thus the number states $\vert1\rangle$, $\vert0\rangle$ are
unable to decay further (as negative occupation numbers are prohibited
by energy conservation). Hence, the matter system decays to the
state in Eq.~\eqref{eq:finalstate-1} where it retains a remnant
coherence $\lambda_{c}$.

Let us now consider two basic examples. We first consider a superposition state of the form: 
\begin{equation} \label{eq:even}
    \vert\psi_{0}\rangle\sim\vert\beta\rangle+\vert-\beta\rangle,
\end{equation}
where $\vert\pm\beta\rangle$ denote coherent states. For $\beta\in\mathbb{R}$ we can interpret $2\beta$ as the superposition size. By writing the state
in number basis one finds that $\vert\psi_{0}\rangle=\sum_{n}\langle2n\vert\psi_{0}\rangle\vert2n\rangle$
which is expected for a state with even parity. The two-phonon process
in Eq.~\eqref{eq:twophonon} will then eventually lead to a decay
of the state $\vert\psi_{0}\rangle$ to the ground state $\vert0\rangle$.

We next consider the superposition of number states 
\begin{equation}\label{eq:simple}
    \vert\psi_{0}\rangle=\frac{1}{\sqrt{2}}\left[\vert0\rangle+\vert1\rangle\right].
\end{equation}
By constructing $\hat{\rho}_{0}=\vert\psi_{0}\rangle\langle\psi_{0}\vert$
we notice we are in the steady state defined in Eq.~\eqref{eq:finalstate-1}
with $\lambda_{0}=\lambda_{1}=\lambda_{c}=1/2$ and the two-phonon process in Eq.~\eqref{eq:twophonon} is energetically forbidden. The state in Eq.~\eqref{eq:simple} or any other superposition of $\vert 0 \rangle$ and  $\vert 1 \rangle$ does not
decay, but rather retains its coherence \emph{indefinitely}.

\subsection{Recovering classical gravitational radiation\label{subsec:Recovering-classical-gravitation}}
To recover the classical results for the gravitational radiation of a harmonic oscillator we will consider coherent states $\vert\beta\rangle$. For coherent states with low occupation number $n=\vert\beta\vert^2 \ll 1$ we have $ \vert\beta\rangle \sim \vert 0 \rangle + \beta \vert 1 \rangle$, which was the case discussed in the previous Sec.~\ref{subsec:Coherence-protection-for-1}. We now consider the  opposite regime $n=\vert\beta\vert^2 \gg 1$ corresponding to coherent states with large occupation numbers. 

For coherent states with large occupation numbers we can replace the operators $\hat{b}$ and $\hat{b}^{\dagger}$  with the corresponding classical observables $b$ and $b^{*}$. From Eq.~\eqref{eq:energydecay} we thus immediately find:
\begin{equation}
\frac{d}{dt}\langle\hat{n}\rangle_{t}\approx-2\gamma_{\text{grav}}\vert b\vert^{4}.\label{eq:energydecay-1}
\end{equation}
We then multiply Eq.~\eqref{eq:energydecay-1} with $\hbar\omega_{\text{m}}$ to obtain: 
\begin{equation}
\dot{E}=-2\hbar\omega_{\text{m}}\gamma_{\text{grav}}\vert b\vert^{4},\label{eq:energydecay-1-1}
\end{equation}
where $E=\hbar\omega_{\text{m}}\langle \hat{n}\rangle_{t}$ denotes the energy of the system.

A high occupation coherent state, neglecting for the moment the energy decay, simply oscillates in a harmonic trap with the position given by $x=\delta_\text{zpf}(b+b^*)$ (where we have again replaced the quantum observables with the corresponding classical ones). The position amplitude is given by $l=2 \delta_\text{zpf}\vert b\vert$, where the zero-point fluctuation $\delta_\text{zpf}$ is defined in Eq.~\eqref{zero-point}. Inverting this relation we thus immediately find:
\begin{equation}
\vert b\vert=l\sqrt{\frac{\mu\omega_{\text{m}}}{2\hbar}}.\label{eq:amplitude-1}
\end{equation}
We now insert in Eq.~\eqref{eq:energydecay-1-1} the amplitude from
Eq.~\eqref{eq:amplitude-1}, the expression for the emission rate
$\gamma_{\text{grav}}=\frac{32}{15}t_{\text{Pl}}^{2}\omega_{\text{m}}^{3}$
from Eq.~\eqref{eq:gammagrav}, and use the definition of the Planck
time $t_{\text{Pl}}=\sqrt{G\hbar/c^{5}}$, to eventually obtain 
\begin{equation}
\dot{E}=-\frac{16GI^{2}\omega_{\text{m}}^{6}}{15c^{5}},\label{eq:energydecay-3}
\end{equation}
where we have introduced the moment of inertia $I=\mu l^{2}$. Importantly,
Eq.~\eqref{eq:energydecay-3} matches exactly the classical linear
quadrupole radiation formula in Eq.~\eqref{eq:edecay}. The obtained
QFT model can be thus seen as the quantum counterpart to the classical theory of
gravitational emission.

\subsection{Coherence protection for the center-of-mass of an isolated system
\label{subsec:Coherence-protection-for}}
We recall that $\hat{\rho}_{t}$ in Eq.~\eqref{eq:lindblad} is the
state associated with the relative motion between two masses with reduced
mass $\mu$ and coupling rate $\omega_{\text{m}}$ (see Fig.~\eqref{fig:quadrupole}). The consequences of the QFT model for the relative motion has been discussed in detail in Secs.~\ref{subsec:energyconservation} - \ref{subsec:Recovering-classical-gravitation}. Here we now analyze the consequences of the QFT model for the \emph{center-of-mass} motion.

We first note that when we decouple the two masses, i.e., $\omega_{\text{m}}\rightarrow0$,
then $\gamma_{\text{grav}}\rightarrow0$, and the decoherence rate
in Eqs.~\eqref{eq:gammagrav} vanishes. This is not surprising, as
the relative motion of two decoupled particles (a system without a
mass quadrupole) is no longer coupled to on-shell gravitation (each
mass will source a gravitational potential, but this happens via exchange
of off-shell gravitons). Importantly, as can be seen from Eqs.~\eqref{eq:harmonic-2}
and \eqref{eq:intlagrangian3} the Hamiltonian for the center-of-mass
motion can be formally mapped to the Hamiltonian of the relative motion
in the limiting case $\omega_{\text{m}}\rightarrow0$ (where the reduced
mass $\mu$ is in place of the total mass $M$). Thus we find that the coherence of 
the \emph{center-of-mass} motion
of an \emph{isolated} system is decoupled from on-shell gravitons
and its coherence will be \emph{completely} protected from quantized gravitational waves.

\section{Summary and discussion} \label{summary}

In this paper, we have developed a quantum field theory (QFT) model
to describe the emission of gravitons from a harmonically trapped
system. The master equation is given in Eq.~\eqref{eq:lindblad}
and the associated decoherence rate $\gamma_\text{grav}$ is given in Eq.~\eqref{eq:gammagrav}.
The key results of the developed QFT model are the following:
\begin{itemize}
 \item conservation of the total energy of the matter systen and gravitational field (Sec.~\ref{subsec:energyconservation}),
 \item decay of number states $\vert n \rangle$ for $n\gg2$ proportional to the square of its corresponding energy $E^2_n$ which can be seen as a consequence of Einstein's equivalence principle and of the quadrupole nature of gravitational waves (Sec.~\ref{subsec:energy decay}),
 \item the quantum harmonic oscillator settles in a steady state with a remnant coherence
of the ground and first excited states as both $\vert 0 \rangle$ and $\vert 1 \rangle$ cannot decay via graviton emission (Sec.~\ref{subsec:Coherence-protection-for-1}),
 \item the formula for classical gravitational radiation is recovered exactly for coherent states with large occupation number (Sec.~\ref{subsec:Recovering-classical-gravitation}),
 \item complete coherence protection for the center-of-mass of
an isolated system which does not have a mass quadrupole and thus cannot emit quantized gravitational waves (Sec.~\ref{subsec:Coherence-protection-for}).
\end{itemize}

Our analysis captures the fact that a harmonic oscillator has a linear mass quadrupole which gives
rise to graviton emission. If however, the mass quadrupole is abstent,
then the system will not emit gravitons. As a result, a free isolated
system or the center-of-mass degree of freedom does not decohere via
graviton emission but rather retains its coherence indefinitely. Only
systems with a mass quadrupole are coupled to on-shell gravitons.
Importantly our analysis recovers the classical gravitational radiation
formulae when we consider coherent states with large occupation numbers
(i.e., classical-like states). 

The developed QFT model predicts a deviation
from classical predictions only when we are close to the ground state
and the quantized nature of the fields becomes important. In particular,
we have found that the state $\sim\vert0\rangle+\vert1\rangle$, or for that matter any
superposition of the ground and first excited states will remain
coherent indefinitely. The reason is that the graviton emission process
only allows transitions $\vert n\rangle\rightarrow\vert n-2\rangle$
for the harmonic oscillator, which can be seen as a consequence
of Einstein's equivalence principle and of the quadrupole nature of
gravitational waves. While linear potentials $\sim x$ vanish by choosing
an inertial reference frame, the quadratic coupling $\sim x^{2}$
cannot be canceled by a change of coordinates and indeed it models
the interaction with ``+'' graviational waves. In the quantum domain,
the quadratic coupling $\hat{x}^{2}$ gives rise to the two-phonon
transitions $\vert n\rangle\rightarrow\vert n-2\rangle$, which is
the only process allowed by energy-momentum conservation.

The decoherence rate $\gamma_\text{grav}\propto \omega_\text{m}^3$ given in Eq.~\eqref{eq:gammagrav} suggested that high-frequency mechanical oscillators could be used for testing the coupling to quantized gravitational waves. 
The analysis showed that the effects can be amplified by using
states with large occupation numbers. However, the QFT
model reduces to the classical predictions for coherent states with large occupation numbers (and hence the quantum and classical
prediction cannot be distinguished experimentally), while other non-trivial quantum states such as superposition states are difficult to achieve experimentally  (see Appendix \ref{sec:QGEM} where we discuss the effects in the next generation of matter-wave interferometry such as the QGEM protocol~\citep{bose2017spin}). Alternatively, we would like
a scheme with a harmonic oscillator near the ground state where quantum
effects become more pronounced, but unfortunately there the graviton
emission process is very slow. We nonetheless hope that the obtained
theoretical results will inspire the development of schemes to test
quantum effects related to quantized gravitational waves such as the
discovered coherence protection mechanism.


\section*{Acknowledgements}

MT acknowledges funding by the Leverhulme Trust (RPG-2020-197).  S.B. would like to acknowledge
EPSRC grants (EP/N031105/1, EP/S000267/1, and EP/X009467/1) and grant
ST/W006227/1.

\appendix

\section{Exact solution\label{sec:time-evolution-with}}

In this appendix, we provide for completeness the solution of the graviton
emission dynamics in Eq.~\eqref{eq:lindblad}. We note that the emission
of two phonons into the gravitational field which can be re-interpreted
as the \emph{absorption of two phonons by the gravitational field}
-- the dynamics can be thus formally mapped to the process of two-photon
absorption by the optical field. We summarize the exact solution following
the presentation of the optical case~\cite {simaan1978off}.

We introduce a normalized time $\tau=2\gamma_{\text{grav}}t$
(such that when $\tau\sim0.5$ we expect to see the first prominent
effects) and the transformed density matrix 
\begin{equation}
\psi_{n}(\mu,\tau)\equiv\sqrt{\frac{(n+\mu)!}{n!}}\langle n\vert\hat{\rho}_{\tau}\vert n+\mu\rangle.
\end{equation}
The solution for the elements with $\mu\neq1$ is given by 
\begin{alignat}{1}
\psi_{n}(\mu,\tau) & =\sum_{\underset{(k-n\,\text{even})}{k=n}}^{\infty}\frac{(-1)^{k/2-n/2}2^{n}}{n!}\nonumber \\
 & \times\frac{\Gamma(k/2+n/2+\sigma)}{\Gamma(\sigma)\Gamma(k/2-n/2+1)}A_{k}^{\sigma}e^{-\lambda_{k}\tau},
\end{alignat}
where 
\begin{alignat}{1}
A_{k}^{\sigma} & =\frac{(k+\sigma)\Gamma(\sigma)}{2^{k}\pi^{1/2}}\sum_{\underset{(m-k\,\text{even})}{m=k}}^{\infty}\frac{m!}{(m-k)}\nonumber \\
 & \times\frac{m!\Gamma(m/2-k/2+1/2)}{(m-k)!\Gamma(m/2+k/2+\sigma+1)}\psi_{m}(\mu,0),
\end{alignat}
$\sigma=\frac{1}{2}(\mu-1)$ and $\lambda_{k}=k(k+\mu-1)+\frac{1}{2}\mu(\mu-1)$.

The solution for elements with $\mu=1$ (with $n>0$) is given by
\begin{alignat}{1}
\psi_{n}(1,\tau) & =\sum_{\underset{(k-n\,\text{even})}{k=n}}^{\infty}\frac{(-1)^{k/2-n/2}2^{n-1}k}{n!}\nonumber \\
 & \times\frac{\Gamma(k/2+n/2)}{\Gamma(k/2-n/2+1)}B_{k}e^{-k^{2}\tau}
\end{alignat}
where 
\begin{alignat}{1}
B_{k} & =\sum_{\underset{(m-k\,\text{even})}{m=k}}^{\infty}\frac{m!}{(m/2+k/2)!(m/2-k/2)!}\nonumber \\
 & \times\frac{1}{2^{2-\delta(k)}}\psi_{m}(1,0),
\end{alignat}
and $\delta(k)=0$ ($\delta(k)=1$) if $k=0$ ($k>0$).

We find that the two-phonon process in Eq.~\eqref{eq:lindblad} induces
a non-trivial steady-state: 
\begin{alignat}{1}
\hat{\rho}_{\infty} & =\lambda_{0}\vert0\rangle\langle0\vert+\lambda_{1}\vert1\rangle\langle1\vert+\lambda_{c}\left[\vert0\rangle\langle1\vert+\vert1\rangle\langle0\vert\right],\label{eq:finalstate}
\end{alignat}
where $\lambda_{0}$,$\lambda_{1}$ and $\lambda_{01}$ depend on
the initial matter state $\hat{\rho}_{0},$and $\vert\,\cdot\,\rangle$
denotes here a number state. In particular, we have that $\lambda_{0}$,$\lambda_{1}$
are the sum of the initial even/odd phonon numbers: 
\begin{alignat}{2}
\lambda_{0} & =\sum_{n=0}^{\infty}\langle2n\vert\hat{\rho}_{0}\vert2n\rangle,\:\: & \lambda_{1} & =\sum_{n=0}^{\infty}\langle2n+1\vert\hat{\rho}_{0}\vert2n+1\rangle,
\end{alignat}
while the steady-state coherence is given by the sum of the initial
coherences between neighboring number states: 
\begin{equation}
\lambda_{c}=\sum_{n=0}^{\infty}\frac{(2n-1)!!}{(2n)!!}\sqrt{\frac{(2n+1)!}{(2n)!}}\langle2n\vert\hat{\rho}_{0}\vert2n+1\rangle.
\end{equation}

\section{QGEM setup} \label{sec:QGEM}
It is interesting to estimate the order of magnitude of gravitational decoherence
for the QGEM (quantum gravity induced entanglement of masses) protocol~\citep{bose2017spin}.

In the original proposal, there are two quantum masses whose center
of mass is separated by a distance $d\sim 450~\text{\ensuremath{\mu}m}$, while their spatial superposition
size is assumed to be $\Delta x\sim250\text{\ensuremath{\mu}m}$. In order to obtain an entanglement
phase of order one -- due to exchange of virtual gravitons -- the
masses (assumed to be the same in the simplest case) were taken to
be $m\sim10^{-14}$ kg. The masses are kept
in a well-preserved vacuum at low temperature to eliminate strong
sources of decoherence mediated via electromagnetic interactions,
and the entire setup is assumed to be in a free fall to minimize the
effect of classical noise sources.

Here we are interested only in a rough upper bound on the gravitational decoherence rate in the QGEM setup. To proceed we make three simplifying approximations. (i) We consider an experiment with only one particle of mass $m$ prepared in a spatial superposition $\Delta x$ (this is reasonable as the two masses are coupled only weakly through gravity and thus the main contribution to decoherence will arise from each particle individually). (ii) Assuming that the interferometric loop is completed in a time $t=1\,\text{s}$ we can estimate an effective harmonic trap frequency as $\omega_{\text{m }}\sim2\pi/t\sim2\pi\times1\,\text{Hz}$ (the particle together with the experimental equipment forms a linear quadrupole during the preparation/recombination of the superposition when the two are coupled by magnetic field gradients).  (iii) The particle delocalization is given by $b=\Delta x/\delta_{\text{zpf}}\sim10^{7}$, where we recall that the zero-point motion of the matter system is $\delta_{\text{zpf}}=\sqrt{\hbar/(2m\omega_{\text{m}})}$. In a more rigorous analysis we would need to decompose the interferometric paths in frequency space~\citep{Toros:2020dbf,wu2023quantum} instead of using an effective harmonic frequency $\omega_{\text{m }}$ and solve the appropriate master equation numerically as well as include the second particle in the modelling,  but the order of magnitude of the effects should not increase.

We now consider the even superposition state given by
\begin{equation}
    \vert\psi \rangle\sim \vert b \rangle +  \vert - b \rangle,
\end{equation}
where we assume $\langle -b\vert b\rangle \approx 0$. 
We know from the analysis below Eq.~\eqref{eq:even} that the state $\vert\psi \rangle$ will eventually decay to the ground state $\vert 0\rangle$. From Eq.~\eqref{eq:energydecay-1} we know that the decay rate of the phonon number for a coherent state $\vert \pm b \rangle$ is given by $\sim\gamma_\text{grav} \vert b\vert^4$. This suggests that a rough upper bound of the decoherence rate could be given by $\gamma_\text{effective}\sim\gamma_\text{grav} \vert b\vert^4$, where $\gamma_\text{grav}\sim t_\text{Pl}^2\omega_\text{m}^3$. 

Plugging in the numbers we find that the order of magnitude of the decoherence rate from graviton emission should not exceed $\gamma_\text{effective}\sim  10^{-56} \text{Hz}$. Decoherence from environmental gravitational waves originating from distant sources would be enhanced by the effective number of gravitons $P t/(\hbar \omega_k)$, where $P$ is the power passing through the interferometer of effective area $A$, $t$ is the interferometric time, and $\omega_k=2\omega_\text{m}$ is the frequency of the gravitons. Assuming $ \Delta x \gg R\sim 1 \mu\text{m}$ (with $R$ denoting the physical size of the particle), we can estimate the effective area to be $A\sim \Delta x R$. However, even if we use $\gamma_\text{effective}\sim\gamma_\text{grav} \vert b\vert^ 4P t/(\hbar \omega_k)$ with $P=10\text{pW}$ (estimated from~\cite{abbott2016observation}) we find only $\gamma_\text{effective}\sim  10^{-33} \text{Hz}$. In short, the QGEM setup does not seem to be prone to be affected by decoherence from quantized gravitational waves.

\bibliographystyle{unsrt}
\bibliography{coherenceprotection}

\begin{thebibliography}{10}

\bibitem{misner1973gravitation}
Charles~W Misner, Kip~S Thorne, John~Archibald Wheeler, et~al.
\newblock {\em Gravitation}.
\newblock Macmillan, 1973.

\bibitem{einstein1916}
Albert Einstein.
\newblock N{\"a}herungsweise integration der feldgleichungen der gravitation.
\newblock {\em Sitzungsberichte der K{\"o}niglich Preu{\ss}ischen Akademie der
  Wissenschaften (Berlin}, page 688, 1916.

\bibitem{einstein1918gravitationswellen}
Albert Einstein.
\newblock {\"U}ber gravitationswellen.
\newblock {\em Sitzungsberichte der K{\"o}niglich Preu{\ss}ischen Akademie der
  Wissenschaften (Berlin}, page 154, 1918.

\bibitem{abbott2016observation}
Benjamin~P Abbott, Richard Abbott, TD~Abbott, MR~Abernathy, Fausto Acernese,
  Kendall Ackley, Carl Adams, Thomas Adams, Paolo Addesso, RX~Adhikari, et~al.
\newblock Observation of gravitational waves from a binary black hole merger.
\newblock {\em Physical review letters}, 116(6):061102, 2016.

\bibitem{yu2020quantum}
Haocun Yu, L~McCuller, M~Tse, N~Kijbunchoo, L~Barsotti, and N~Mavalvala.
\newblock Quantum correlations between light and the kilogram-mass mirrors of
  ligo.
\newblock {\em Nature}, 583(7814):43--47, 2020.

\bibitem{abbott2009observation}
B~Abbott, R~Abbott, R~Adhikari, P~Ajith, Bruce Allen, G~Allen, R~Amin,
  SB~Anderson, WG~Anderson, MA~Arain, et~al.
\newblock Observation of a kilogram-scale oscillator near its quantum ground
  state.
\newblock {\em New Journal of Physics}, 11(7):073032, 2009.

\bibitem{whittle2021approaching}
Chris Whittle, Evan~D Hall, Sheila Dwyer, Nergis Mavalvala, Vivishek Sudhir,
  R~Abbott, A~Ananyeva, C~Austin, L~Barsotti, J~Betzwieser, et~al.
\newblock Approaching the motional ground state of a 10-kg object.
\newblock {\em Science}, 372(6548):1333--1336, 2021.

\bibitem{calzetta1994noise}
Esteban Calzetta and BL~Hu.
\newblock Noise and fluctuations in semiclassical gravity.
\newblock {\em Physical Review D}, 49(12):6636, 1994.

\bibitem{anastopoulos1996quantum}
C~Anastopoulos.
\newblock Quantum theory of nonrelativistic particles interacting with gravity.
\newblock {\em Physical Review D}, 54(2):1600, 1996.

\bibitem{anastopoulos2013master}
C~Anastopoulos and BL~Hu.
\newblock A master equation for gravitational decoherence: probing the textures
  of spacetime.
\newblock {\em Classical and Quantum Gravity}, 30(16):165007, 2013.

\bibitem{riedel2013evidence}
C~Jess Riedel.
\newblock Evidence for gravitons from decoherence by bremsstrahlung.
\newblock {\em arXiv preprint arXiv:1310.6347}, 2013.

\bibitem{blencowe2013effective}
MP~Blencowe.
\newblock Effective field theory approach to gravitationally induced
  decoherence.
\newblock {\em Physical review letters}, 111(2):021302, 2013.

\bibitem{suzuki2015environmental}
Fumika Suzuki and Friedemann Queisser.
\newblock Environmental gravitational decoherence and a tensor noise model.
\newblock In {\em Journal of Physics: Conference Series}, volume 626, page
  012039. IOP Publishing, 2015.

\bibitem{de2015decoherence}
VA~De~Lorenci and LH~Ford.
\newblock Decoherence induced by long wavelength gravitons.
\newblock {\em Physical Review D}, 91(4):044038, 2015.

\bibitem{oniga2016quantum}
Teodora Oniga and Charles H-T Wang.
\newblock Quantum gravitational decoherence of light and matter.
\newblock {\em Physical Review D}, 93(4):044027, 2016.

\bibitem{oniga2017quantum}
Teodora Oniga and Charles H-T Wang.
\newblock Quantum coherence, radiance, and resistance of gravitational systems.
\newblock {\em Physical Review D}, 96(8):084014, 2017.

\bibitem{quinones2017quantum}
Diego~A Qui{\~n}ones, Teodora Oniga, Benjamin~TH Varcoe, and Charles H-T Wang.
\newblock Quantum principle of sensing gravitational waves: From the zero-point
  fluctuations to the cosmological stochastic background of spacetime.
\newblock {\em Physical Review D}, 96(4):044018, 2017.

\bibitem{vedral2020decoherence}
Vlatko Vedral.
\newblock Decoherence of massive superpositions induced by coupling to a
  quantized gravitational field.
\newblock {\em arXiv preprint arXiv:2005.14596}, 2020.

\bibitem{xu2020toy}
Qidong Xu and MP~Blencowe.
\newblock Toy models for gravitational and scalar qed decoherence.
\newblock {\em arXiv preprint arXiv:2005.02554}, 2020.

\bibitem{parikh2020noise}
Maulik Parikh, Frank Wilczek, and George Zahariade.
\newblock The noise of gravitons.
\newblock {\em International Journal of Modern Physics D}, page 2042001, 2020.

\bibitem{parikh2020quantum}
Maulik Parikh, Frank Wilczek, and George Zahariade.
\newblock Quantum mechanics of gravitational waves.
\newblock {\em arXiv preprint arXiv:2010.08205}, 2020.

\bibitem{parikh2020signatures}
Maulik Parikh, Frank Wilczek, and George Zahariade.
\newblock Signatures of the quantization of gravity at gravitational wave
  detectors.
\newblock {\em arXiv preprint arXiv:2010.08208}, 2020.

\bibitem{bose2017spin}
Sougato Bose, Anupam Mazumdar, Gavin~W Morley, Hendrik Ulbricht, Marko
  Toro{\v{s}}, Mauro Paternostro, Andrew~A Geraci, Peter~F Barker, MS~Kim, and
  Gerard Milburn.
\newblock Spin entanglement witness for quantum gravity.
\newblock {\em Physical review letters}, 119(24):240401, 2017.

\bibitem{ICTS}
\url{https://www.youtube.com/watch?v=0Fv-0k13s_k}, 2016.
\newblock Accessed 1/11/22.

\bibitem{marletto2017gravitationally}
Chiara Marletto and Vlatko Vedral.
\newblock Gravitationally induced entanglement between two massive particles is
  sufficient evidence of quantum effects in gravity.
\newblock {\em Physical review letters}, 119(24):240402, 2017.

\bibitem{Marshman:2019sne}
Ryan~J. Marshman, Anupam Mazumdar, and Sougato Bose.
\newblock {Locality \& Entanglement in Table-Top Testing of the Quantum Nature
  of Linearized Gravity}.
\newblock {\em Phys. Rev. A}, 101(5):052110, 2020.

\bibitem{bose2022mechanism}
Sougato Bose, Anupam Mazumdar, Martine Schut, and Marko Toro{\v{s}}.
\newblock Mechanism for the quantum natured gravitons to entangle masses.
\newblock {\em Physical Review D}, 105(10):106028, 2022.

\bibitem{Vinckers:2023grv}
Ulrich K.~Beckering Vinckers, \'Alvaro de~la Cruz-Dombriz, and Anupam Mazumdar.
\newblock {Quantum entanglement of masses with nonlocal gravitational
  interaction}.
\newblock {\em Phys. Rev. D}, 107(12):124036, 2023.

\bibitem{Carney_2019}
Daniel Carney, Philip~CE Stamp, and Jacob~M Taylor.
\newblock Tabletop experiments for quantum gravity: a user's manual.
\newblock {\em Classical and Quantum Gravity}, 36(3):034001, 2019.

\bibitem{Carney:2021vvt}
Daniel Carney.
\newblock {Newton, entanglement, and the graviton}.
\newblock {\em Phys. Rev. D}, 105(2):024029, 2022.

\bibitem{danielson2022gravitationally}
Daine~L Danielson, Gautam Satishchandran, and Robert~M Wald.
\newblock Gravitationally mediated entanglement: Newtonian field versus
  gravitons.
\newblock {\em Physical Review D}, 105(8):086001, 2022.

\bibitem{christodoulou2023locally}
Marios Christodoulou, Andrea Di~Biagio, Markus Aspelmeyer, {\v{C}}aslav
  Brukner, Carlo Rovelli, and Richard Howl.
\newblock Locally mediated entanglement in linearized quantum gravity.
\newblock {\em Physical Review Letters}, 130(10):100202, 2023.

\bibitem{Elahi:2023ozf}
Shafaq~Gulzar Elahi and Anupam Mazumdar.
\newblock Probing massless and massive gravitons via entanglement in a warped
  extra dimension.
\newblock {\em Physical Review D}, 108(3), August 2023.

\bibitem{Biswas:2022qto}
Dripto Biswas, Sougato Bose, Anupam Mazumdar, and Marko Toro\v{s}.
\newblock {Gravitational Optomechanics: Photon-Matter Entanglement via Graviton
  Exchange}, 9 2022.

\bibitem{Bose:2022czr}
Sougato Bose, Anupam Mazumdar, Martine Schut, and Marko Toro\v{s}.
\newblock {Entanglement Witness for the Weak Equivalence Principle}.
\newblock {\em Entropy}, 25(3):448, 2023.

\bibitem{hanif2023testing}
Farhan Hanif, Debarshi Das, Jonathan Halliwell, Dipankar Home, Anupam Mazumdar,
  Hendrik Ulbricht, and Sougato Bose.
\newblock Testing whether gravity acts as a quantum entity when measured.
\newblock {\em arXiv preprint arXiv:2307.08133}, 2023.

\bibitem{gorini1976completely}
Vittorio Gorini, Andrzej Kossakowski, and Ennackal Chandy~George Sudarshan.
\newblock Completely positive dynamical semigroups of n-level systems.
\newblock {\em Journal of Mathematical Physics}, 17(5):821--825, 1976.

\bibitem{lindblad1976generators}
Goran Lindblad.
\newblock On the generators of quantum dynamical semigroups.
\newblock {\em Communications in Mathematical Physics}, 48(2):119--130, 1976.

\bibitem{simaan1978off}
HD~Simaan and R~Loudon.
\newblock Off-diagonal density matrix for single-beam two-photon absorbed
  light.
\newblock {\em Journal of Physics A: Mathematical and General}, 11(2):435,
  1978.

\bibitem{gilles1993two}
L~Gilles and PL~Knight.
\newblock Two-photon absorption and nonclassical states of light.
\newblock {\em Physical Review A}, 48(2):1582, 1993.

\bibitem{de1986introduction}
Venzo De~Sabbata and Maurizio Gasperini.
\newblock {\em Introduction to gravitation}.
\newblock World Scientific Publishing Company, 1986.

\bibitem{maggiore2008gravitational}
Michele Maggiore.
\newblock {\em Gravitational waves: Volume 1: Theory and experiments},
  volume~1.
\newblock Oxford university press, 2008.

\bibitem{Burgess:2003jk}
C.P. Burgess.
\newblock {Quantum gravity in everyday life: General relativity as an effective
  field theory}.
\newblock {\em Living Rev. Rel.}, 7:5--56, 2004.

\bibitem{poisson2011motion}
Eric Poisson, Adam Pound, and Ian Vega.
\newblock The motion of point particles in curved spacetime.
\newblock {\em Living Reviews in Relativity}, 14(1):7, 2011.

\bibitem{will2006confrontation}
Clifford~M Will.
\newblock The confrontation between general relativity and experiment.
\newblock {\em Living reviews in relativity}, 9(1):3, 2006.

\bibitem{visser2018post}
Matt Visser.
\newblock Post-newtonian particle physics in curved spacetime.
\newblock {\em arXiv preprint arXiv:1802.00651}, 2018.

\bibitem{rakhmanov2014fermi}
Malik Rakhmanov.
\newblock Fermi-normal, optical, and wave-synchronous coordinates for spacetime
  with a plane gravitational wave.
\newblock {\em Classical and Quantum Gravity}, 31(8):085006, 2014.

\bibitem{Weinberg:1972kfs}
Steven Weinberg.
\newblock {\em {Gravitation and Cosmology}: {Principles and Applications of the
  General Theory of Relativity}}.
\newblock John Wiley and Sons, New York, 1972.

\bibitem{breuer2002theory}
Heinz-Peter Breuer, Francesco Petruccione, et~al.
\newblock {\em The theory of open quantum systems}.
\newblock Oxford University Press on Demand, 2002.

\bibitem{louisell1970quantum}
WH~Louisell.
\newblock Quantum optics.
\newblock {\em Academic, New York, 1969) pp}, 680:742, 1970.

\bibitem{gardiner2004quantum}
Crispin Gardiner and Peter Zoller.
\newblock {\em Quantum noise: a handbook of Markovian and non-Markovian quantum
  stochastic methods with applications to quantum optics}.
\newblock Springer Science \& Business Media, 2004.

\bibitem{schlosshauer2007decoherence}
Maximilian~A Schlosshauer.
\newblock {\em Decoherence: and the quantum-to-classical transition}.
\newblock Springer Science \& Business Media, 2007.

\bibitem{agarwal2012quantum}
Girish~S Agarwal.
\newblock {\em Quantum optics}.
\newblock Cambridge University Press, 2012.

\bibitem{Toros:2020dbf}
Marko Toro{\v{s}}, Thomas~W Van De~Kamp, Ryan~J Marshman, MS~Kim, Anupam
  Mazumdar, and Sougato Bose.
\newblock Relative acceleration noise mitigation for nanocrystal matter-wave
  interferometry: Applications to entangling masses via quantum gravity.
\newblock {\em Physical Review Research}, 3(2):023178, 2021.

\bibitem{wu2023quantum}
Meng-Zhi Wu, Marko Toro{\v{s}}, Sougato Bose, and Anupam Mazumdar.
\newblock Quantum gravitational sensor for space debris.
\newblock {\em Physical Review D}, 107(10):104053, 2023.

\end{thebibliography}

\end{document}